\documentclass[runningheads]{llncs}
\usepackage{graphicx}
\usepackage[table,xcdraw]{xcolor}
\usepackage{tablefootnote}
\usepackage{fancyhdr}
\begin{document}
\pagestyle{fancy}
\fancyhf{}
\fancyhead[LE,LO]{Accepted for publication in the 15th International Symposium on Visual Computing (ISVC 2020)}
\title{
Ink Marker Segmentation in Histopathology Images Using Deep Learning
\thanks{This work was supported by an NSERC-CRD Grant on ``Design and Development of Devices and Procedures for Recognizing Artefacts and Foreign
Tissue Origin for Diagnostic Pathology''}}

\author{Danial Maleki\inst{1} \and Mehdi Afshari\inst{1} \and
Morteza Babaie\inst{1} \orcidID{0000-0002-6916-5941} \and H.R. Tizhoosh\inst{1} \orcidID{0000-0001-5488-601X} }
\authorrunning{F. Author et al.}
%
\institute{Kimia Lab, University of Waterloo, Canada \\ (dmaleki, m4afshar, mbabaie and tizhoosh)@uwaterloo.ca }


%
%
%
\maketitle              
\begin{abstract}
Due to the recent advancements in machine vision, digital pathology has gained significant attention. Histopathology images are distinctly rich in visual information. The tissue glass slide images are utilized for disease diagnosis. Researchers study many methods to process histopathology images and facilitate fast and reliable diagnosis; therefore, the availability of high-quality slides becomes paramount. The quality of the images can be negatively affected when the glass slides are ink-marked by pathologists to delineate regions of interest. As an example, in one of the  largest public histopathology datasets, The Cancer Genome Atlas (TCGA), approximately $12\%$ of the digitized slides are affected by manual delineations through ink markings. To process these open-access slide images and other repositories for the design and validation of new methods, an algorithm to detect the marked regions of the images is essential to avoid confusing tissue pixels with ink-colored pixels for  computer methods. In this study, we propose to segment the ink-marked areas of pathology patches through a deep network. A dataset from $79$ whole slide images with $4,305$ patches was created and different networks were trained. Finally, the results showed an FPN model with the EffiecentNet-B3 as the backbone was found to be the superior configuration with an F1 score of $94.53\%$.
\end{abstract}
\section{Introduction}
The definitive diagnosis of numerous diseases is possible through meticulous visual inspection by a trained pathologist, an examination that requires considerable time and effort while inevitably being prone to error. Computer vision and machine learning can assist the pathologists to reduce their workload and to increase the diagnostic accuracy \cite{fuchs2011computational}. Computer-assisted diagnosis necessitates the digitization of biopsy glass slides. In some cases, pathologists use ink markers to highlight parts of glass slides for a variety of purposes such as educational and diagnosis hints. Once the slides are digitized, marker signs naturally appear in the digital images as well. Consequently, computer methods are potentially prone to mistaking marker colored regions for tissue information \cite{ali2019ink}.

The pathology slides contain valuable information in high resolution; therefore, rejection of an entire slide because of some ink markings may not be a feasible approach. To enable reliable processing with images with marker signs, we have used deep learning models to detect and removed different marker colors. Several deep learning strategies have been explored to generate a mask that extracts the ink marker areas automatically. The aim of this task is to distinguish between the tissue areas and the areas that are colorized through the marker's ink. Elimination of areas that have marker signs is useful since in most cases areas that are covered by marker are of less importance (therefore overpainted by the pathologists to show something more important). As a result, discarding these areas may generally not affect the tissue relevant for diagnosis. Moreover, removing manual markings can help to generate a considerable number of proper patches without the presence of the marker artifacts for the benefit of many machine learning algorithms.

Smart software for digital scanners is  another potential application of this research. The focus depth of whole slide scanners must be adjusted for different tissue regions due to variable tissue thickness. Focus points on the marker area will considerably affect the scan quality. Avoiding to set focus points on the marker places from a pre-scan image, could boost the automated scanning process and hence improve the lab workflow in digital pathology \cite{babaie2019deep}.

Three main challenges concerning the detection of ink-marking areas should be addressed. The first one is that these markings can be drawn by pens with different colors. The worst case in this regard occurs when ink markers have a color similar to the tissue such as red or pink. The second problem is the composition of inconsistent patterns and ink markers' shapes/symbols. They can be letters, circles, or other shapes such as arrows and dots. The third issue is that ink markers can be created/seen at different transparency levels. In addition, markers may cross the tissue regions, mark outside or around the tissue, which can affect the observed color of the marker. A combination of these scenarios in a single image makes the discovery of marked regions a challenging task  \cite{venkatesh2019restoration}. 


Traditional methods, such as thresholding, may not be accurate enough because of the difficulties mentioned above. An automatic approach that is capable of overcoming these issues is the goal of this study. The advent of deep learning has led to multiple breakthroughs in different domains. One of these domains is medical imaging which consists of numerous tasks including classification, image search, and segmentation  \cite{Babaie_2017_CVPR_Workshops,8489574}. In this work, a comparative study of deep network segmentation models (U-net and FPN) that can generate a mask of areas that contain ink markers is presented. The areas that include ink marker are determined, and then a binary mask of the detected regions is generated. By having this mask, no trace of markers may appear in the patch selection phase avoiding  the loss of tissue information. The method is accurate, tolerant and requires only a limited number of images for training. The final network as well as the created dataset are available to download for academic purposes.

\section{Related Work}

Ink marker segmentation could be considered as a part of the pre-processing of histopathology images. The main goal of pre-processing methods is to prepare high-qulaity images for the training procedure. Most of the pre-processing methods for histopathology images ignore the presence of ink marker in images. Taqi et al. \cite{taqi2018review} investigated various types of artifacts that may appear in histopathological images and how could one  differentiate an artifact and a tissue segement. The presence of these artifacts may render images useless for computer aided diagnosis when accessing existing repositories. Rastogi et al.  \cite{rastogi2013artefacts} demonstrated that the presence of artifact can be a major pitfall. The artifacts may occur in different stages of the process. Therefore, it is essential to identify the presence of artifacts. Janowczyk et al.  \cite{janowczyk2017stain} mentioned that variability of histopathology images can mislead the diagnostic algorithms. Stain normalization using sparse autoencoders has been proposed to avoid the problem.  However, the presence of ink markers as an artifact has not been discussed in these studies.


A few techniques are developed for the detection and removal of the ink marker signs in histopathology images. These methods could be categorized into two following types. The first type is the image processing methods based on traditional algorithms. Most of the techniques that are proposed to remove marker signs use thresholding techniques combined with mathematical morphology operations. In this approach, different types of color space, filters, and thresholds for different ink marker colors have been used\footnote{https://github.com/deroneriksson/python-wsi-preprocessing/blob/master/docs/wsi-preprocessing-in-python/index.md}. 
HistoQC is a tool presented for the quality control of WSIs which discovers artifacts and outputs outliers. The method was validated using open-access dataset of TCGA and the results were verfied by pathologists. However, the precision of the outlier detection of the method is not high enough which can result in loss of valuable data \cite{janowczyk2019histoqc}.
Mostly, these methods are not fully automated. Due to the high variation of the markers' color and intensity, thresholds fail in different images. As a result, the manual setting of the thresholds or other parameters for each image may be required.

The second method type is deep learning. Recent improvements in deep learning enable the modeling of complex data distributions. One of the well-known tasks that are done using deep learning techniques is image  segmentation and reconstruction.  Venkatesh et al. \cite{venkatesh2019restoration} used CycleGan for the reconstruction of a marker removed WSI. Sharib et al.  \cite{ali2019ink} separated tiles that were contaminated with markers with a binary classification network and fed the tiles with marker signs to a YOLO3 network  \cite{redmon2018yolov3}  to detect the marker areas. Finally, for the restoration of the original non-marked tiles, CycleGAN was used \cite{zhu2017unpaired}.

\subsection{Segmentation models}
\begin{figure}
    \centering
    \includegraphics[width=0.9\linewidth]{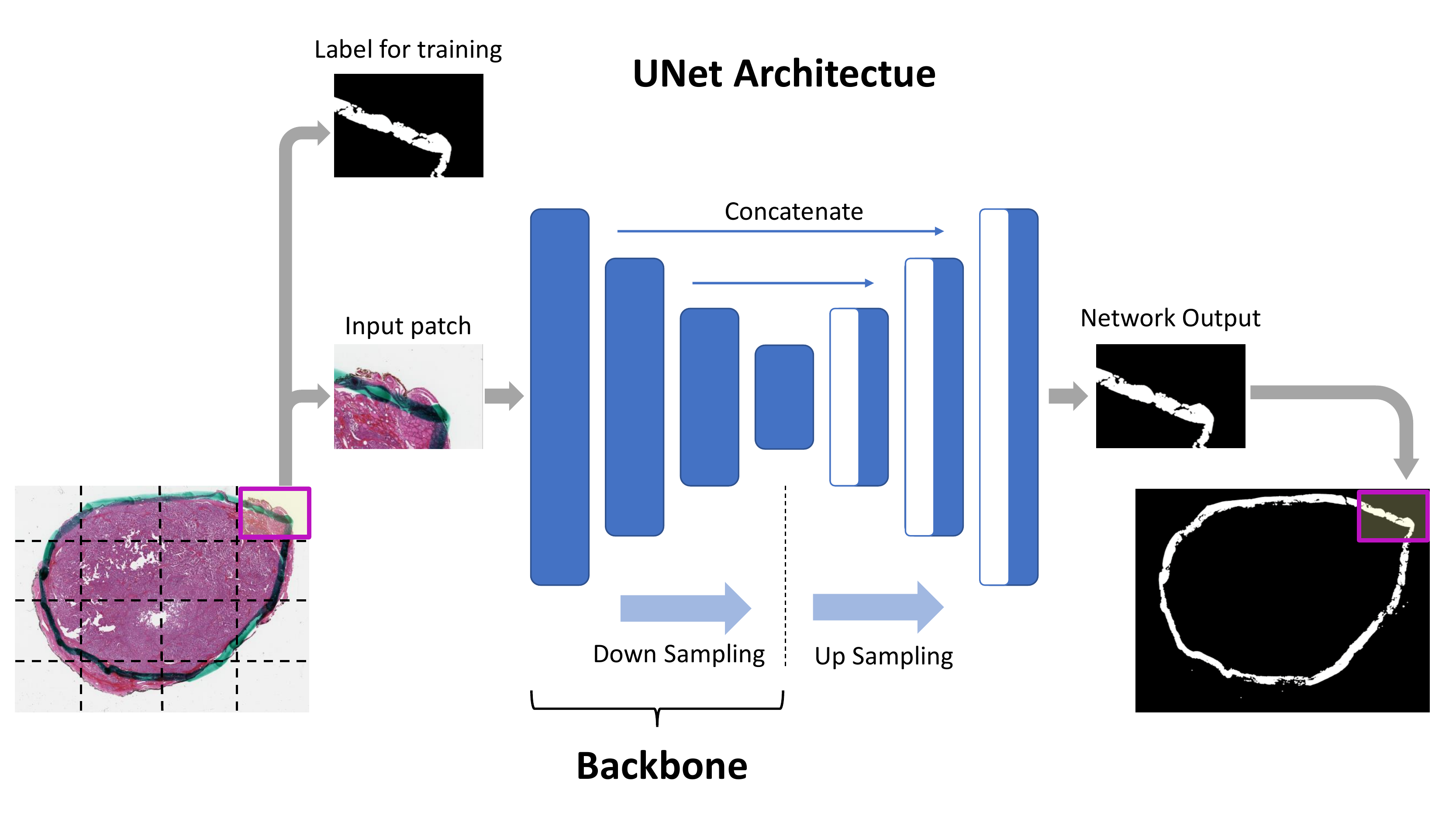}
    \caption{Overall structure of the proposed method for segmentation of ink marked images using U-Net architecture network. Blue boxes show the CNN layers.}
    \label{model_unet}
\end{figure}

U-Net has been developed to segment  biomedical images \cite{ronneberger2015u}. The network architecture has two major parts. The first part is the encoding path which is used to capture the image context. In this part, convolutional layers are applied which followed by max-pooling layers. The second part uses transposed convolution layers and up-sampling layers to construct an image with the same size as the input image. The main goal is to add skip connections from the down-sampling part to the second part as an up-sampling part which can help the model to have better localization information. The down-sampling layers output is then concatenated with the input of up-sampling layers. Figure \ref{model_unet} shows the  proposed configuration when U-Net architecture is used. For the pathology practice, U-Net has been utilized for a wide range of applications, for instance  for epithelial tissue  segmentation \cite{bulten2018automated} and cell nuclei segmentation \cite{naylor2018segmentation}.

\begin{figure}
    \centering
    \includegraphics[width=0.9\linewidth]{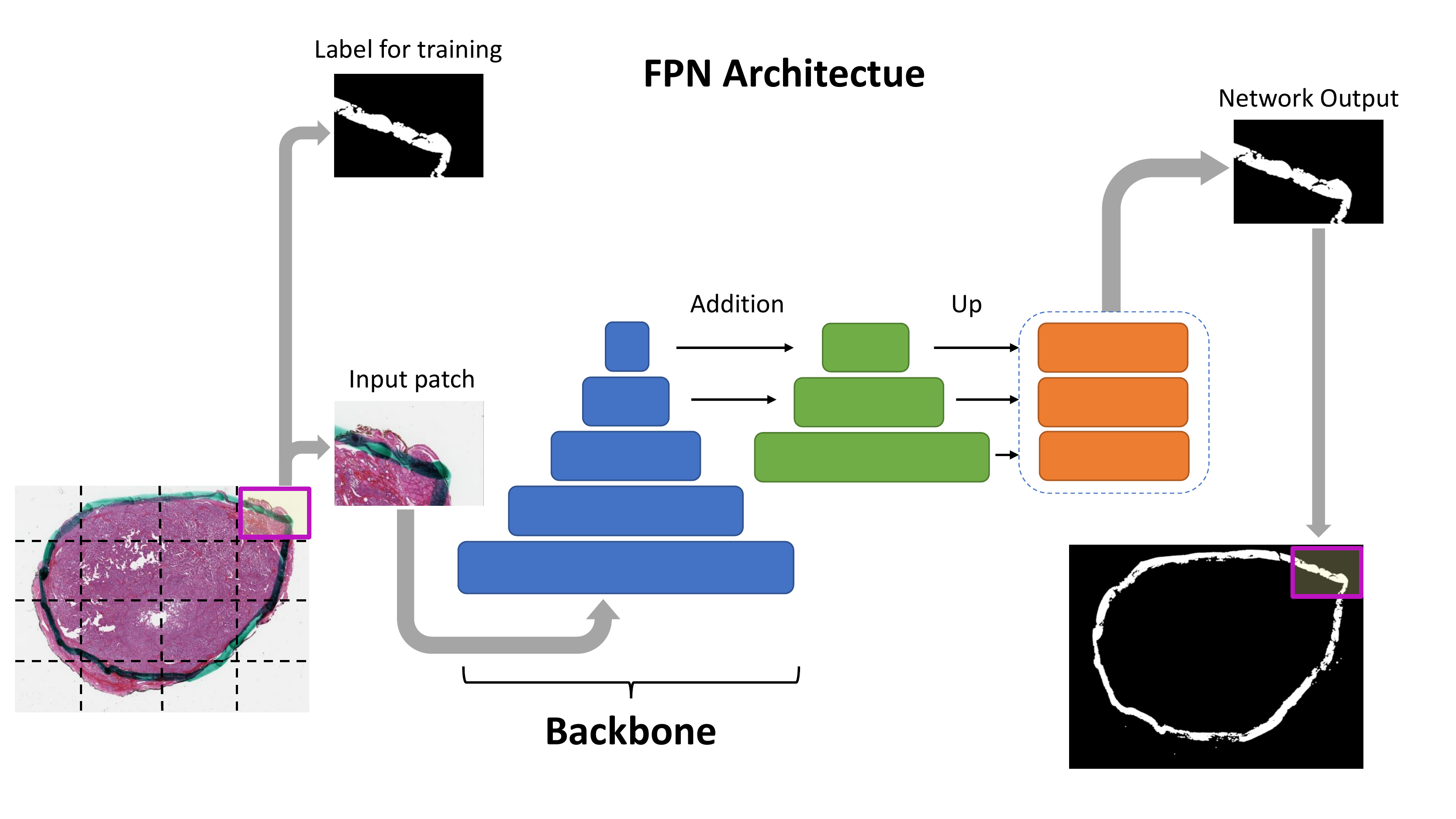}
    \caption{The proposed architecture (blue/green/orange boxes show the CNN layers).}
    \label{model_fpn}
\end{figure}

Feature Pyramid Network (FPN) \cite{lin2017feature} like U-Net has two parts. The first part which is a bottom-up pathway has  a convolutional network as  backbone. The output of each stage in this part will be used as a skip connection for enriching the second part layers. The second part of the model, which is named  top-down pathway, uses transposed convolution and up-sampling layers to up-sample features. Afterward, FPN concatenates each stage in this pathway with skip connections from the bottom-up pathway. In the end, the network concatenates the output of each stage after feeding each of them to the convolutional layer with a kernel size of three and predicts the segmentation for the input image.  Fig. \ref{model_fpn} shows the FPN architecture used in this study.

\subsection{Transfer learning}
Providing large sets of labeled data in medical imaging is expensive \cite{kalra2019pan}. Recently, transfer learning has become a feasible in the medical imaging domain \cite{kieffer2017convolutional}. Transfer learning helps to build new models in a shorter time requiring a rather small number of training samples. In computer vision, transfer learning is mainly implemented through the use of pre-trained models. A pre-trained model is a network that is trained on a large dataset for a similar task. A pre-trained model which was trained on the ImageNet dataset  \cite{deng2009imagenet} with more than 14 million images for the encoding pathway that can extract features by using a few images from a new domain for the segmentation task. For instance, in EffiecientNet \cite{tan2019efficientnet} the main idea is to find an optimal baseline network using Neural Architecture Search (NAS) with respect to a trade-off between accuracy and FLOPS (Floating-point Operations Per Second) and then scale the depth, width and resolution through an equation. By changing the parameter in the formula, different network settings can be achieved that have different FLOPS and accuracy levels.
 
 In ResNet \cite{he2016deep}, using skip connections is the main idea. Skip connections between ResNet blocks help the network to perform the backward propagation of the loss better and to prevent gradient vanishing that is a common problem in very deep networks.
 
 \section{Method}
The goal of this study is a fully automatic Convolutional Neural Network (CNN) that generates a binary mask with the same size of the input image. 
Two of the most popular architectures for the medical image segmentation namely U-Net and FPN, are chosen and investigated in this paper. These networks achieved success in many segmentation tasks mainly due to skipping connections from the down-sampling pathways to the up-sampling pathways. On the other hand, due to the limited  number of training samples, transfer learning approach is utilized by using a pre-trained network as the backbone of segmentation models. 
Different backbone networks are compared to discover a suitable pre-trained network for the segmentation task. Selected backbones are EfficientNet and ResNet.


In this study, our network uses a combination of two types of cost functions. One of them is the Dice loss function and the other one is the Focal loss function  \cite{lin2017focal}. The dice loss function principally measures the overlap between two binary samples. The measurement range is 0 to 1 where a dice coefficient value of 1 denotes a perfect match between two samples which in this case would be the target and the network output. The dice loss function can be calculated as follows:

\begin{equation}
    \textrm{Dice}= \frac{2|Output \cap Target|}{|Output|+|Target|},
\end{equation}

where $|output \cap target|$ denotes the common pixels between output and target and $|output|$ shows the number of pixels in output (and likewise for target).

The second term of the cost function is the focal loss. The distribution of each class pixels in the data is imbalanced which serves the definition of focal loss term. The term avoids the complication of correct pixels number and  the total number of pixels. Relatively speaking, uncomplicated examples are well classified; therefore, these examples contribute to the loss rather in a minor fashion. The focal loss function can be described as:

\begin{equation}
    F L\left(p_{t}\right)=-\left(1-p_{t}\right)^{\gamma} \log \left(p_{t}\right).
\end{equation}

In this formula, $\gamma$ is a tunable parameter. This controlling element defines the steepness of the function shape. 
Parameter $p_t$ is the probability of the background class. The term puts more emphasis on hard samples while reduces loss for  well-classified examples.

\section{Experiments and Results} 

\begin{figure}[ht]
    \centering
    \includegraphics[width=0.49\textwidth]{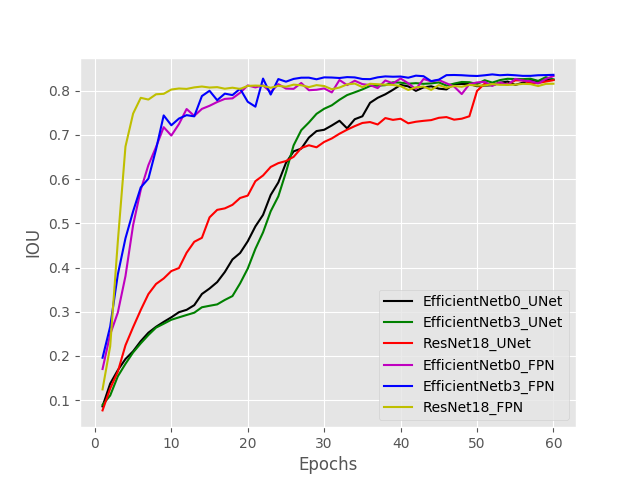}
    \includegraphics[width=0.49\textwidth]{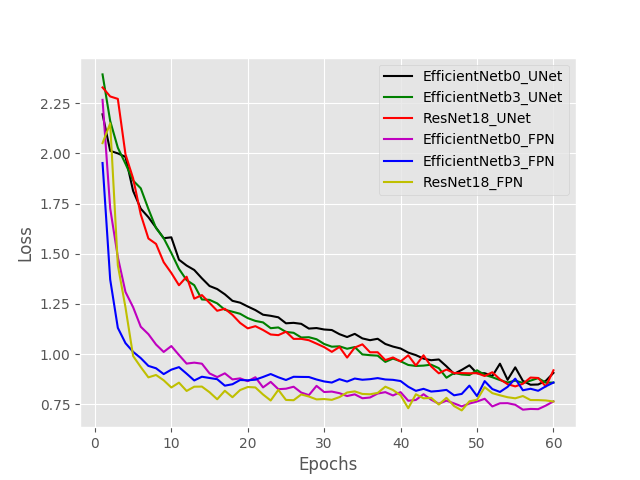}
    \caption{Jaccard index (left) and loss function values (right) for validation data.}
    \label{iou}
\end{figure}

To train the model, 89 ink marked TCGA \cite{tomczak2015cancer} slides were selected visually through diverse body parts. The ground truth masks (ink marker areas) of selected slides are created manually. Then, each WSI has been patched into images with the size of $256 \times 256$ pixels at $1\times$ magnification. The total number of 4,305 patches are extracted from 89 annotated slides to train the model. We applied 8-fold cross validation to train and test our model. Therefore, training, validation, and test data are set to $70\%$, $15\%$ and $15\%$ from the randomly selected patches, respectively. The encoder pathway is the pre-trained model and different models are trained using the Adam optimizer  \cite{kingma2014adam} with the learning rate set to 0.00001. Each model is trained for 60 epochs with one Nvidia Tesla V100 GPU. The maximum training time of a network  is roughly 160 seconds per epoch.
On the other hand, the minimum training time is 60 seconds per epoch for the smallest size network as U-Net architecture and EfficientNet-B0 backbone. Two evaluation criteria are used. The first one is the Jaccard index defined as:

\begin{equation}
    \textrm{Jaccard index} =\frac{|Output \cap Target|}{|Output \cup Target|}.
\end{equation}

The second one is the F1-score which measures the sensitivity and specificity together. 
As discussed in the method section, the proposed model can adapt to different architectures and backbones. FPN and U-Net are used for the architecture and EfficienNet-B0, EfficeinetNet-B3 and ResNet-18 are used for the encoder part.  Fig. \ref{iou}(left) shows the Jaccard index for all training scenarios over the training process. The vertical axis shows trained epochs and the horizontal axis shows the Jaccard index for the validation data. The Jaccard index trend shows the fastest learning trend of FPN architecture with ResNet backbone with respect to number of epochs. Our study shows that ResNet-18 with FPN architecture has the steepest slope at the starting epochs, but at the end, other backbones show better performance. UNet architecture with EfficientNet-B0 and EfficientNet-B3 show similar trends as the slowest learners with respect to the number of training epochs. All networks reach a considerably high Jaccard index at the end of training process. Comparison among the validation loss values are shown in Fig. \ref{iou} (right) for all architectures and backbones. As expected, the loss value over epochs has correlation with Jaccard  criterion. Fig. \ref{iou} (left) shows the importance of architecture based on loss. Different architectures show varied behaviour in the course of loss value trends. FPN architecture has lower loss value in comparison with UNet architecture through all epochs. 
  

Table \ref{tab:my-table} shows the evaluation on the test set (average of 8-fold cross validations) of the trained models on the presented dataset. In the ink marker segmentation problem, FPN architecture shows better performance comparing U-Net architecture. FPN architecture with ResNet-18 backbone performed less accurate compared with other networks. The reason of lower F1-score is the lower capacity of the network. However, the smaller size of network results in favorable computation time. The results  shows that FPN architecture is a fast learner for the desired task compared with the U-Net architecture. The reason of this faster learning trend is the additional part that appears after $1\times 1$ convolution that FPN architecture has. 

\begin{figure}[ht]
    \centering
    \includegraphics[width=0.9\textwidth]{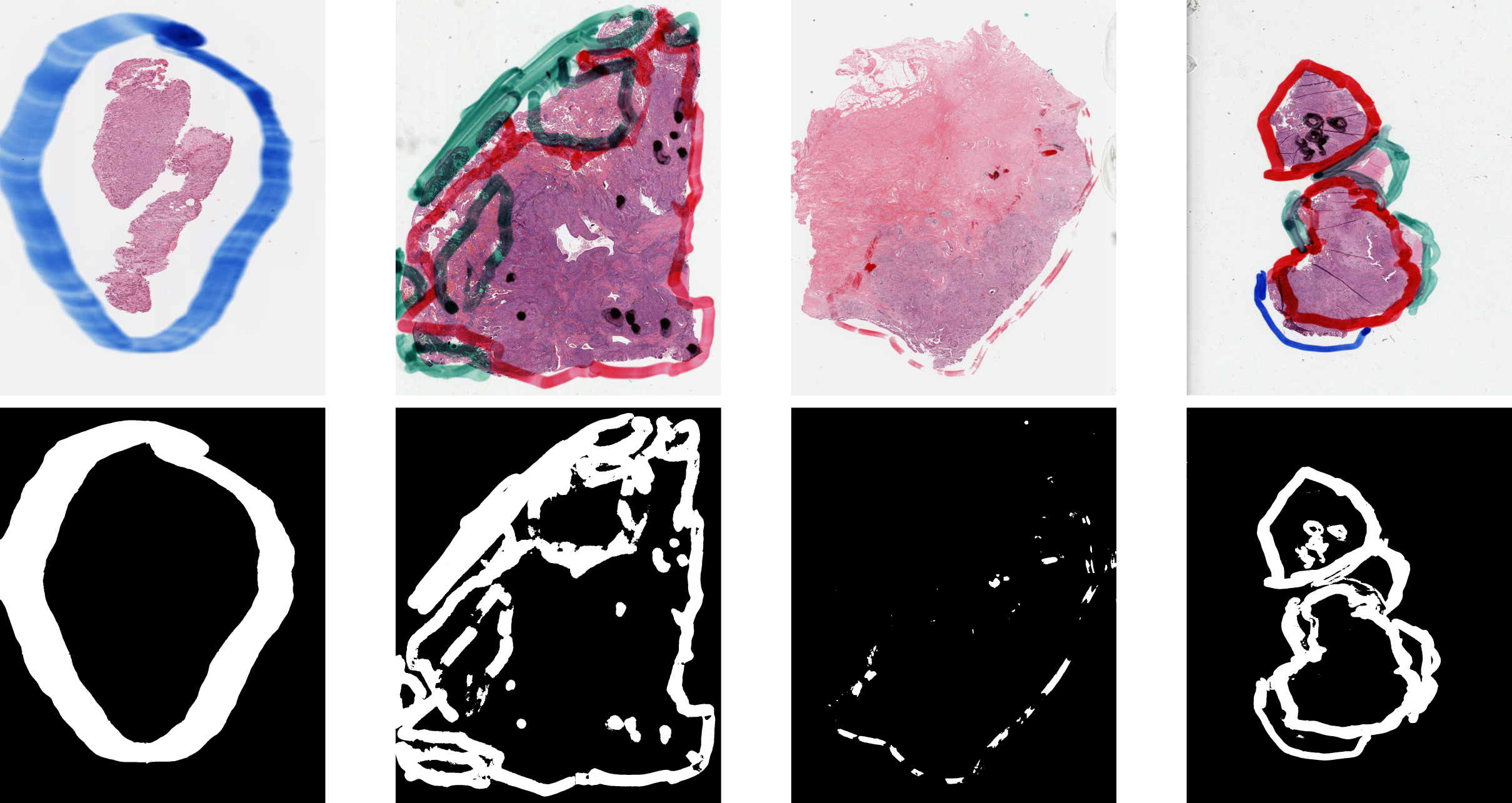}
    \caption{WSIs (top) and ground-truths (bottom): The network segmentation is applied on individual patches and then combined to a complete image.}
    \label{inoutpair}
\end{figure}

Fig. \ref{inoutpair} shows some sample input-output pairs of trained networks. The samples include different colors, intensities, and overlaps with the tissue.

\begin{table}[ht]
\centering
\caption{Evaluation of results for Jaccard index and F1-score and execution times}
\label{tab:my-table}
\begin{tabular}{lc|c|c|c}
\cline{3-4}
                                            & \multicolumn{1}{l|}{}                     & \multicolumn{2}{c|}{\cellcolor[HTML]{FFFFFF}\textbf{Metrics}} & \multicolumn{1}{l}{\cellcolor[HTML]{FFFFFF}{\color[HTML]{333333} \textbf{}}} \\ \hline
\multicolumn{1}{|c|}{\textbf{Base}} & \cellcolor[HTML]{FFFFFF}\textbf{Backbone} & \multicolumn{1}{l|}{IoU}    & \multicolumn{1}{l|}{F1 Score}   & \multicolumn{1}{l|}{Time (s)}                                          \\ \hline \hline
\multicolumn{1}{|l|}{}                      & \cellcolor[HTML]{FFFFFF}EfficientNet-B0    & 0.8244                      & 0.9128                        & \multicolumn{1}{c|}{0.0238}                                                   \\ \cline{2-5} 
\multicolumn{1}{|c|}{U-Net}                  & \cellcolor[HTML]{FFFFFF}EfficientNet-B3    & 0.8251                      & 0.9227                        & \multicolumn{1}{c|}{0.029}                                                    \\ \cline{2-5} 
\multicolumn{1}{|l|}{}                      & \cellcolor[HTML]{FFFFFF}ResNet-18          & 0.8234                      & 0.9091                          & \multicolumn{1}{c|}{0.0195}                                                   \\ \hline \hline
\multicolumn{1}{|l|}{}                      & \cellcolor[HTML]{FFFFFF}EfficientNet-B0    & 0.8326                        & 0.9312                         & \multicolumn{1}{c|}{0.028}                                                   \\ \cline{2-5} 
\multicolumn{1}{|c|}{FPN}                   & \cellcolor[HTML]{FFFFFF}EfficientNet-B3    & 0.8354                      & 0.9453                          & \multicolumn{1}{c|}{0.0352}                                                   \\ \cline{2-5} 
\multicolumn{1}{|l|}{}                      & \cellcolor[HTML]{FFFFFF}ResNet-18          & 0.8154                  & 0.8912                          & \multicolumn{1}{c|}{0.0217}                                                    \\ \hline \hline
\multicolumn{1}{|c|}{Baseline\tablefootnote{https://github.com/deroneriksson/python-wsi-preprocessing/blob/master/docs/wsi-preprocessing-in-python/index.md}}                   & \cellcolor[HTML]{FFFFFF}-    & 0.6512                      & 0.7312                          &  \multicolumn{1}{c|}{0.0189}                                                  \\ \hline
\end{tabular}
\end{table}

\section{Conclusions}

A fully automatic deep model to generate a binary mask of areas that have ink markers on archived WSIs is proposed and evaluated in this study. The investigated method is capable of the extraction of ink markers with different colors and different shapes and intensities. Calculation of two evaluation metrics including the Jaccard index and F1-score showed the efficiency of the method. Removing ink markings helps to use all archived WSIs that have ink highlights for research instead of discarding them. As well, this method can be added to digital scanners to apply the ink removal during the image acquisition.

\bibliographystyle{splncs04}
\bibliography{Bibliography}





\end{document}